# Compression of Far-Fields in the Fast Multipole Method via Tucker Decomposition

Cheng Qian, Mingyu Wang, and Abdulkadir C. Yucel, *Senior Member, IEEE*

*Abstract*— **Tucker decomposition is proposed to reduce the memory requirement of the far-fields in the fast multipole method (FMM)-accelerated surface integral equation simulators. It is particularly used to compress the far-fields of FMM groups, which are stored in three-dimensional (3-D) arrays (or tensors). The compressed tensors are then used to perform fast tensor-vector multiplications during the aggregation and disaggregation stages of the FMM. For many practical scenarios, the proposed Tucker decomposition yields a significant reduction in the far-fields' memory requirement while dramatically accelerating the aggregation and disaggregation stages. For the electromagnetic scattering analysis of a $30\lambda$-diameter sphere, it reduces the memory requirement of the far-fields more than 87% while it expedites the aggregation and disaggregation stages by a factor of 15.8 and 15.2, respectively, where $\lambda$ is the wavelength in free space.**

## I. INTRODUCTION

The fast multipole method (FMM)-accelerated surface integral equation (SIE) simulators have been widely used for many applications ranging from the composite and plasmonic nanostructure analysis to wireless channel characterization [1]-[4]. These simulators' applications to large-scale problems are often limited due to their large memory requirements, primarily stemming from three data structures. These data structures include the near-field interaction matrices, the data structures holding the translation operator samples, and the matrices storing the far-fields of FMM groups. So far, substantial research has been performed to reduce the memory requirements of the data structures involving the near-field interactions and translation operators [5]-[10]. This study focuses on the memory saving in the data structures storing the far-field signatures of the FMM groups.

In past, several techniques have been proposed to reduce the memory requirement of far-fields of FMM groups. Truncated singular value decomposition (SVD) [11] was applied to the compression of the matrices holding the far-fields (and receiving patterns) of FMM groups (i.e., aggregation and disaggregation matrices). This technique yielded a significant reduction in the memory requirement of the far-fields. In [3], SVD was applied to the far-fields of individual basis function in each FMM group. The SVD technique was then leveraged in a multilevel FMM (MLFMM) scheme [12]. Apart from the SVD technique, the aggregation and disaggregation matrices were compressed by a skeleton basis in a hierarchical MLFMM algorithm [10]. All these methods proposed to reduce the far-fields' memory requirement are highly valuable and matrix decomposition-based. However, there is still room for improvement as the tensor decomposition methods often exhibit higher compression performance compared to the matrix decomposition methods [13, 14].

In this study, Tucker decompositions for compressing far-fields are proposed. During the simulator's setup stage, the far-fields of the FMM groups are computed and stored in three-dimensional arrays (or tensors), which are then compressed by the Tucker decompositions. During the simulator's iterative solution stage, the aggregation and disaggregation stages of the FMM are performed using the Tucker-compressed tensors with a substantially reduced computational cost. The numerical results show that the Tucker decompositions yield a significant reduction in far-fields' memory and computational time requirement in aggregation and disaggregation stages. Bearing the advantages of the proposed tensor methodology in mind, this paper's contribution can be summarized as follows. First, it introduces a tensor decomposition methodology, Tucker decomposition, for compressing the far-fields. Although tensor decompositions have recently been applied to integral equation simulators [15]-[19], this paper presents the application of a tensor decomposition methodology to the far-fields for the first time. Second, it provides the algorithms for performing rapid tensor-vector multiplications with minimal cost during the aggregation and disaggregation stages. Third, it demonstrates the performance of the Tucker decomposition while the simulation parameters are varied. These simulation parameters include the FMM box size, decomposition tolerance, and FMM accuracy. Note that the studies in [7, 8] also leverage Tucker decompositions but not for compressing the far-fields and not for performing the operations in Tucker format in a fast way, as done in this study.

Our numerical results show that the proposed Tucker

Manuscript received November 23, 2020. This work was supported by Ministry of Education, Singapore, under grant AcRF TIER 1-2018-T1-002-077 (RG 176/18), and the Nanyang Technological University under a Start-Up Grant. (*Corresponding author: Abdulkadir C. Yucel.*)

Cheng Qian, Mingyu Wang, and Abdulkadir C. Yucel are with the School of Electrical and Electronic Engineering, Nanyang Technological University, Singapore 639798 (e-mail: cqian@ntu.edu.sg; mingyu003@e.ntu.edu.sg; acyucel@ntu.edu.sg).

decomposition outperforms the traditional SVD technique. For the electromagnetic (EM) scattering analysis of a 30 $\lambda$-diameter sphere, the proposed Tucker decomposition reduces the far-fields' memory requirement by a factor of 7.8 while the traditional SVD technique [11] can reduce that by a factor of 3.9, when the FMM box size and decomposition tolerance is set to 2 $\lambda$ and $10^{-6}$. Here $\lambda$ is the wavelength. For the same problem, the proposed Tucker decomposition allows accelerating the aggregation and disaggregation stages by a factor of 15.8 and 15.2, while the SVD technique expedites those stages by a factor of 8.3 and 8.6, respectively. The memory saving achieved by the proposed Tucker decomposition increases with increasing FMM box size (and consequently increasing number of basis functions in the FMM group), FMM accuracy and decomposition tolerance.

The rest of this paper is organized as follows. Section II briefly explains the FMM-accelerated SIE simulator; here the fast Fourier transform enhanced FMM (a.k.a FMM-FFT) is leveraged to explain the compressed far-fields and accelerated aggregation and disaggregation stages. Section II then introduces the Tucker decomposition for the tensorized far-fields and algorithms for fast tensor-vector multiplications performed during the aggregation and disaggregation stages. Section III presents the numerical results illustrating the compression and acceleration achieved by the proposed Tucker decompositions. Finally, the conclusions are drawn in Section IV.

## II. FORMULATION

### A. FMM-Accelerated SIE Simulator

In this subsection, the FMM-accelerated SIE simulator's formulation for the analysis of EM scattering from perfectly electric conducting (PEC) surfaces is expounded. In this analysis, an arbitrarily-shaped PEC surface $S$ with unit normal $\hat{\mathbf{n}}$ resides in free-space with permittivity $\varepsilon_0$ and permeability $\mu_0$. The surface is excited by an incident E-field $\mathbf{E}^{inc}(\mathbf{r})$ at frequency $f$. The incident E-field induces surface current $\mathbf{J}$, which in turn generates the scattered E-field $\mathbf{E}^{sca}(\mathbf{r}, \mathbf{J})$. By invoking the boundary conditions on $S$, the electric field SIE is written as

$$\hat{\mathbf{n}} \times \hat{\mathbf{n}} \times \mathbf{E}^{inc}(\mathbf{r}) = -\hat{\mathbf{n}} \times \hat{\mathbf{n}} \times \mathbf{E}^{sca}(\mathbf{r}, \mathbf{J})$$
$$= \hat{\mathbf{n}} \times \hat{\mathbf{n}} \times j\omega\mu_0 \int_S \left(1 + \frac{\nabla\nabla'}{k^2}\cdot\right) \mathbf{J}(\mathbf{r}') G(\mathbf{r}, \mathbf{r}') d\mathbf{r}'. \quad (1)$$

Here $G(\mathbf{r}, \mathbf{r}') = \exp(-jk|\mathbf{r} - \mathbf{r}'|)/(4\pi|\mathbf{r} - \mathbf{r}'|)$ is the Green's function, $\omega = 2\pi f$, $k = \omega\sqrt{\mu_0\varepsilon_0}$, $\mathbf{r}$ and $\mathbf{r}'$ denote the observation and source locations, respectively. To solve (1), $\mathbf{J}$ is discretized using the Rao-Wilton-Glisson (RWG) basis functions $\mathbf{b}_n(\mathbf{r})$ [20] as $\mathbf{J}(\mathbf{r}) = \sum_{n=1}^{N} \mathbf{b}_n(\mathbf{r}) I_n$. Substituting the discretized current into (1) and applying Galerkin testing (with $\mathbf{b}_m(\mathbf{r})$, $m = 1, \dots, N$) yields a linear system of equations (LSE), which reads as

$$\mathbf{V} = \bar{\mathbf{Z}} \cdot \mathbf{I}. \quad (2)$$

Here $\bar{\mathbf{Z}}$ is the method of moments matrix with entries $\bar{\mathbf{Z}}_{m,n} = \langle \mathbf{b}_m, -\hat{\mathbf{n}} \times \hat{\mathbf{n}} \times \mathbf{E}^s(\mathbf{r}, \mathbf{b}_n) \rangle$, $\mathbf{V}$ is the excitation vector with entries $\mathbf{V}_m = \langle \mathbf{b}_m, \hat{\mathbf{n}} \times \hat{\mathbf{n}} \times \mathbf{E}^i(\mathbf{r}) \rangle$, and $\mathbf{I}$ is the current coefficient vector with entries $\mathbf{I}_n = I_n$, where $\langle \cdot, \cdot \rangle$ denotes the inner product. The iterative solution of LSE in (2) requires $O(N^2)$ CPU and memory cost. This cost can be reduced to $O(N^{3/2})$ via a single-level FMM [21], $O(N^{4/3} \log^{2/3} N)$ via the FMM-FFT [22], or $O(N \log N)$ via the MLFMM [23].

In the FMM-FFT scheme, the computational domain enclosing $S$ is partitioned into $K_x$, $K_y$, and $K_z$ boxes along $x$, $y$, and $z$ directions, respectively; $K = K_x K_y K_z$. The boxes with centers $\mathbf{r}_v$ reside on a structured grid and are labeled by $B_v$, where $v = (v_x, v_y, v_z)$ is the multi-index with $v_x = 1, \dots, K_x$, $v_y = 1, \dots, K_y$, and $v_z = 1, \dots, K_z$. Typically in this scheme, the interactions between basis functions in two non-empty boxes (i.e., groups) touching each other via a shared corner, edge, or surface are computed classically. On the other hand, the interactions between basis functions in far groups, not touching each other, are computed via aggregation, translation, and disaggregation stages (see [7, 8] for details). Here we just provide a brief description of the aggregation and disaggregation stages to show the data structures being compressed, which are not explicitly shown in [7, 8]; the explanation of the translation stage can be found in [7, 8]. In the aggregation stage, far-fields of basis functions in a group are multiplied by the corresponding current coefficients as

$$\mathcal{F}_\mathbf{u}^\alpha = \bar{\mathbf{P}}_\mathbf{v}^{\alpha,+} \cdot \mathbf{I}_\mathbf{v} \quad (3)$$

where $\mathcal{F}_\mathbf{u}^\alpha$ denotes the outgoing wave spectra tensor with dimensions $K_x \times K_y \times K_z \times N_{dir}$, defined for $\alpha = \{\theta, \varphi\}$ direction, $\bar{\mathbf{P}}_\mathbf{v}^{\alpha,+}$ is the aggregation matrix of group $B_\mathbf{v}$ with dimensions $N_{dir} \times N_\mathbf{v}$, and $\mathbf{I}_\mathbf{v}$ is a vector with $N_\mathbf{v}$ entries, storing the current coefficients corresponding to the basis functions in $B_\mathbf{v}$, where $N_\mathbf{v}$ and $N_{dir}$ are the numbers of the basis functions in group and plane-wave directions, respectively. The subscript $\mathbf{u} = (v_x, v_y, v_z, i)$ represents a multi-index with $v_x = 1, \dots, K_x$, $v_y = 1, \dots, K_y$, $v_z = 1, \dots, K_z$, and $i = 1, \dots, N_{dir}$. The entries of $\mathcal{F}_\mathbf{u}^\alpha$ corresponding to empty boxes are zero. Each column of the aggregation matrix stores the far-fields of each basis function, i.e., $\bar{\mathbf{P}}_\mathbf{v}^{\alpha,+} = [\mathbf{P}_1^{\alpha,+} \dots \mathbf{P}_n^{\alpha,+} \dots \mathbf{P}_{N_\mathbf{v}}^{\alpha,+}]$, $n \in B_\mathbf{v}$, where the entries of the vector $\mathbf{P}_n^{\alpha,+}$ are computed via

$$P_{n,i}^{\alpha,\pm} = \hat{\alpha} \cdot \int_{S_n} (\bar{\mathbf{I}} - \hat{\mathbf{k}}_i \hat{\mathbf{k}}_i) \mathbf{b}_n(\mathbf{r}) \exp(\pm jk\hat{\mathbf{k}}_i \cdot (\mathbf{r} - \mathbf{r}_\mathbf{v})) d\mathbf{r} \quad (4)$$

$n \in B_\mathbf{v}$, $i = 1, \dots, N_{dir}$. Here $S_n$ is the support of basis function $\mathbf{b}_n(\mathbf{r})$. $\hat{\mathbf{k}}_i$, $i = 1, \dots, N_{dir}$, denotes the plane-wave directions pointing towards the spherical grid points, obtained by the Cartesian product of $L+1$ points selected along $\theta$-direction and $2L+1$ points selected along $\varphi$-direction (See [24] for the details on the spherical grid obtained by quadrature rules). $N_{\text{dir}}$ is assigned to $(L+1)(2L+1)$, $L$ is computed via $L=2kR^s+(2kR^s)^{1/3}1.8(\log_{10}(\Gamma^{-1}))^{2/3}$ [25], where $\Gamma$ denotes the desired number of accurate digits in FMM approximation and $R^s$ is the radius of the smallest sphere that can enclose the groups. In (4), $\hat{\boldsymbol{\alpha}}=\{\hat{\boldsymbol{\theta}},\hat{\boldsymbol{\varphi}}\}$ and only $\theta$ and $\phi$ components of the patterns are considered as $\bar{\mathbf{I}}-\hat{\mathbf{k}}_i\hat{\mathbf{k}}_i=\hat{\boldsymbol{\theta}}\hat{\boldsymbol{\theta}}+\hat{\boldsymbol{\varphi}}\hat{\boldsymbol{\varphi}}$. The aggregation matrices $\bar{\mathbf{P}}_{\mathbf{v}'}^{\alpha,+}$ are computed and stored during the setup stage of the simulator. Those are compressed by the Tucker decomposition, as explained in Section II.B. In the disaggregation stage, the resulting incoming plane-wave spectra is projected onto each basis function $\mathbf{b}_m$ in $B_{\mathbf{v}'}$ via

$$\mathbf{V}_{\mathbf{v}'}^- = \sum_\alpha \bar{\mathbf{P}}_{\mathbf{v}'}^{\alpha,-} \mathcal{G}_{\mathbf{u}'}^\alpha. \qquad (5)$$

Here, $\mathcal{G}_{\mathbf{u}'}^\alpha$ stores incoming plane-wave spectra for all plane-wave directions, $\bar{\mathbf{P}}_{\mathbf{v}'}^{\alpha,-}$ denotes the disaggregation matrix of $B_{\mathbf{v}'}$ with dimensions $N_{\mathbf{v}'}\times N_{\text{dir}}$. $\mathbf{V}_{\mathbf{v}'}^-$ stands for a vector with $N_{\mathbf{v}'}$ entries, storing the contributions to the matrix-vector product for the $N_{\mathbf{v}'}$ basis functions in $B_{\mathbf{v}'}$. Each row of the disaggregation matrix stores the receiving pattern of each basis function, i.e., $\bar{\mathbf{P}}_{\mathbf{v}'}^{\alpha,-}=[\mathbf{P}_1^{\alpha,-}\ldots\mathbf{P}_m^{\alpha,-}\ldots\mathbf{P}_{N_{\mathbf{v}'}}^{\alpha,-}]^T$, $m\in B_{\mathbf{v}'}$. In general, the entries of $\mathbf{P}_m^{\alpha,-}$ are computed via (4). That said, the $\mathbf{P}_m^{\alpha,-}$ can be directly obtained from $\mathbf{P}_m^{\alpha,+}$ via complex conjugation for the EM analysis of scatterers residing in free-space.

### B. Tucker Decomposition

For the sake of simplicity, consider the aggregation matrix of a group, i.e., $\bar{\mathbf{P}}_{\mathbf{v}}^{\alpha,+}$ of $B_{\mathbf{v}}$, with dimensions $N_{\text{dir}}\times N_{\mathbf{v}}$ and the disaggregation matrix of the same group, $\bar{\mathbf{P}}_{\mathbf{v}}^{\alpha,-}=(\bar{\mathbf{P}}_{\mathbf{v}}^{\alpha,+})^*$, where $*$ denotes the conjugate transpose operation. Let $\bar{\mathbf{A}}$ represent the aggregation matrix $\bar{\mathbf{P}}_{\mathbf{v}}^{\alpha,+}$ and let $n_\theta n_\varphi$ and $n_\beta$ correspond to $N_{\text{dir}}$ and $N_{\mathbf{v}}$, respectively. Here $n_\theta=L+1$ and $n_\varphi=2L+1$ are the numbers of quadrature points selected along $\theta$- and $\varphi$-directions on the spherical grid, respectively. The aggregation matrix $\bar{\mathbf{A}}$ with dimensions $n_\theta n_\varphi\times n_\beta$ can be converted to a tensor $\mathcal{A}$ with dimensions $n_\theta\times n_\varphi\times n_\beta$ [Fig. 1(a)]. This tensor, called 'aggregation tensor' here, is naturally low-rank since $\bar{\mathbf{A}}$ is rank-deficient and compressible by SVD as shown in [11]. Henceforth the aggregation tensor $\mathcal{A}$ is also compressible via its Tucker decomposition, which reads as [Fig. 1(b)]

$$\mathcal{A} = \mathcal{C}_{\text{T}} \times_1 \bar{\mathbf{U}}_{\text{T}}^\theta \times_2 \bar{\mathbf{U}}_{\text{T}}^\varphi \times_3 \bar{\mathbf{U}}_{\text{T}}^\beta, \qquad (6)$$

where $\mathcal{C}_{\text{T}}$ denotes the core tensor with dimensions $r_\theta\times r_\varphi\times r_\beta$, $\bar{\mathbf{U}}_{\text{T}}^p$ represents a factor matrix with dimensions $n_p\times r_p$,

$p=\{\theta,\varphi,\beta\}$, and $\times_q$, $q=\{1,2,3\}$ denotes the mode $-q$ matrix product. As the multilinear ranks $r_p$ are much smaller than $n_p$, $p=\{\theta,\varphi,\beta\}$, and $\prod_p r_p + \sum_p n_p r_p << \prod_p n_p$, the Tucker compressed representation of $\mathcal{A}$ requires much less memory compared to $\mathcal{A}$. The detailed procedure for obtaining core tensor and factor matrices by utilizing the high-order singular value decomposition [26] is provided in [7, 8] and briefly described here. To compute $\bar{\mathbf{U}}_{\text{T}}^\theta$ for example, first, the unfolding matrix of $\mathcal{A}$ pertinent to the first dimension ($\theta$-direction) is obtained by reordering the elements in $\mathcal{A}$. Such reordering yields an unfolding matrix with dimensions $n_\theta\times n_\varphi n_\beta$, which has the first dimension overlapping with the first dimension of the tensor $\mathcal{A}$ with dimensions $n_\theta\times n_\varphi\times n_\beta$. Next, the SVD of the unfolding matrix is efficiently computed and truncated by considering a prescribed tolerance $\gamma$. The truncated unitary matrix of SVD is assigned as the factor matrix $\bar{\mathbf{U}}_{\text{T}}^\theta$. The same procedure is applied for the second and third dimensions. Using the resulting factor matrices, $\bar{\mathbf{U}}_{\text{T}}^\theta$, $\bar{\mathbf{U}}_{\text{T}}^\varphi$, and $\bar{\mathbf{U}}_{\text{T}}^\beta$, the core tensor is obtained via $\mathcal{C}_{\text{T}} = \mathcal{A} \times_1 \bar{\mathbf{U}}_{\text{T}}^{\theta*} \times_2 \bar{\mathbf{U}}_{\text{T}}^{\varphi*} \times_3 \bar{\mathbf{U}}_{\text{T}}^{\beta*}$.

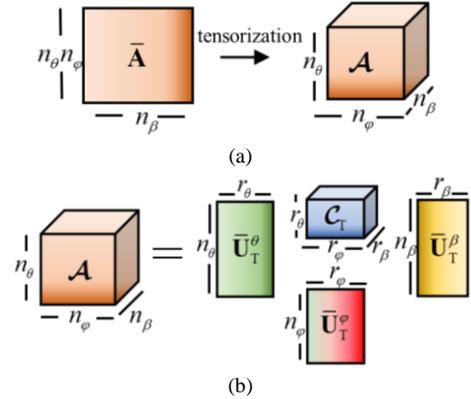

Fig. 1. (a) Tensorization of the aggregation matrix $\bar{\mathbf{A}}$. (b) Conceptual representation of Tucker decomposition of aggregation tensor $\mathcal{A}$. (Note: In conceptual representation, each edge of each shape corresponds to one dimension. The dimensions of 1-D, 2-D, 3-D, 4-D arrays are / will be represented on the edges of a line, rectangle, cube, and trapezoidal, respectively).

### C. Tensor-Vector Multiplication in Compressed Format

After obtaining Tucker-compressed representations (i.e., core tensors and factor matrices) of the aggregation tensors for all groups during the setup stage, those are used to perform the aggregation and disaggregation stages rapidly [(3) and (5)] during the solvers' iterative solution stage. The aggregation is traditionally performed for each FMM group by multiplying the aggregation matrix $\bar{\mathbf{A}}$ with a temporary current coefficient vector $\mathbf{I}^-$ [Fig. 2(a)]. This matrix-vector multiplication is accelerated using the Tucker-compressed representation of $\mathcal{A}$, as explained in the procedure in Fig. 2(b). In the first step of this procedure, the core tensor $\mathcal{C}_{\text{T}}$ is reshaped into a matrix with dimensions $r_\theta \times r_\theta r_\varphi$, which is multiplied with the result of the product of $\bar{\mathbf{U}}_T^\beta$ and $(\mathbf{I}^-)^T$. In step 2, the resulting vector with dimensions $1 \times r_\theta r_\varphi$ is reshaped into a matrix with dimensions $r_\varphi \times r_\theta$, which is multiplied with $\bar{\mathbf{U}}_T^\varphi$. In step 3, the transpose of the resulting matrix is multiplied with $\bar{\mathbf{U}}_T^\theta$. In the final step, the resulting matrix is converted to a vector and stored as a mode-4 fiber of $\mathcal{F}_u^\alpha$.

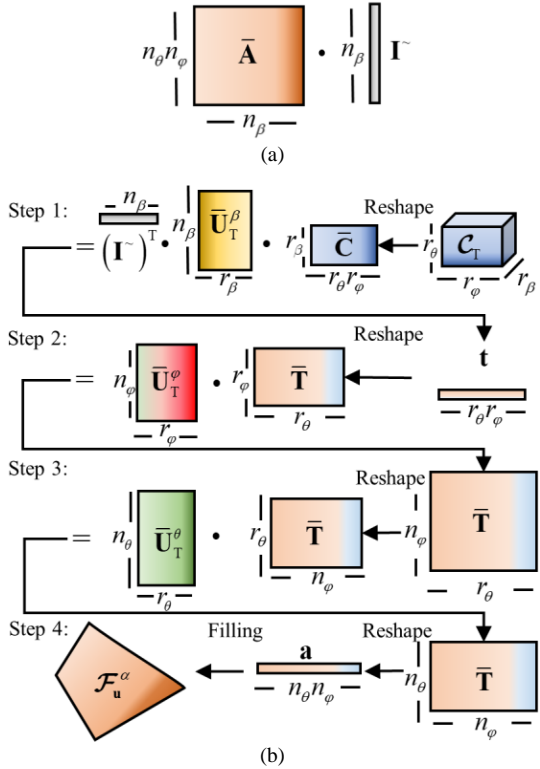

Fig. 2. Aggregation stage: (a) Multiplication of aggregation matrix $\bar{\mathbf{A}}$ with a temporary current coefficient vector $\mathbf{I}^-$. (b) The procedure for fast tensor-vector multiplication in Tucker-compressed representation for aggregation.

Similarly, the Tucker-compressed representation can be used in the disaggregation stage directly as the disaggregation matrix is the conjugate transpose of the aggregation matrix. During the disaggregation stage, the disaggregation matrix $\bar{\mathbf{A}}^*$ is traditionally multiplied by a temporary vector $\mathbf{g}$, storing the group's incoming wave spectra [Fig. 3(a)]. To perform this stage fast, the Tucker-compressed representation is multiplied by $\mathbf{g}$ via the procedure outlined in Fig. 3(b). In the first step of this procedure, the conjugate of $\mathbf{g}$ is reshaped into a matrix with dimensions $n_\theta \times n_\varphi$ and multiplied by the transpose of $\bar{\mathbf{U}}_T^\theta$. The resulting matrix is then used in the matrix-matrix multiplication with $\bar{\mathbf{U}}_T^\varphi$. In step 2, the resulting matrix is vectorized and multiplied by the matricized core tensor $\mathcal{C}_T$. In step 3, the resulting vector is multiplied with $\bar{\mathbf{U}}_T^\beta$. The conjugate of the resulting vector holds 'far' contributions to the basis functions in the group.

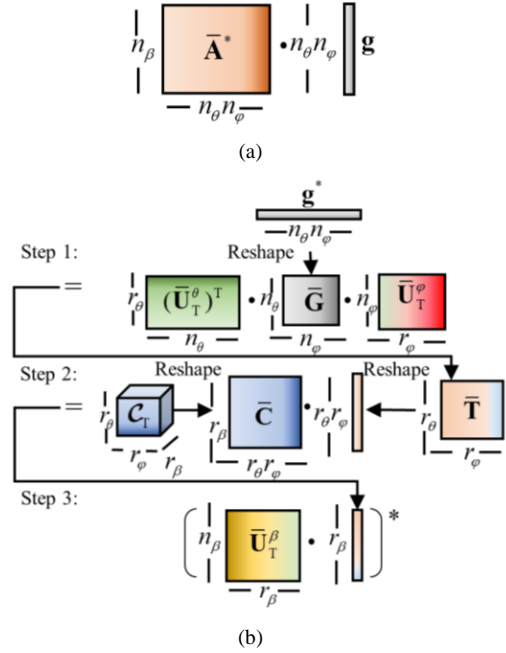

Fig. 3. Disaggregation stage: (a) Multiplication of disaggregation matrix $\bar{\mathbf{A}}^*$ with a temporary vector $\mathbf{g}$. (b) The procedure for fast tensor-vector multiplication via Tucker-compressed representation for disaggregation.

For the aggregation and disaggregation stages, the conventional truncated SVD method requires $O(n_\beta r_\beta + n_\theta n_\varphi r_\beta)$ operations for each FMM group; here $r_\beta$ is the rank of the SVD. Compared to the truncated SVD, which can be considered as the subset Tucker (a.k.a., high-order SVD), Tucker scheme further compresses along $\theta$ and $\varphi$ directions (in addition to $\beta$ direction), which introduces additional reduction in the memory and computational cost. The computational cost of performing aggregation and disaggregation stages via Tucker-compressed representation is $O(n_\beta r_\beta + n_\theta n_\varphi r_\theta + n_\varphi r_\varphi r_\theta + r_\beta r_\varphi r_\theta)$. The memory requirements of SVD and Tucker schemes are $O(n_\beta r_\beta + n_\theta n_\varphi r_\beta)$ and $O(n_\beta r_\beta + n_\theta r_\varphi + n_\theta r_\theta + r_\beta r_\varphi r_\theta)$, respectively. Since $r_\theta << n_\theta$, $r_\varphi << n_\varphi$, and $r_\theta << r_\beta$ the computational cost and memory requirement of Tucker scheme are much smaller than those of the SVD scheme.

## III. NUMERICAL RESULTS

This section presents several numerical examples demonstrating the memory and computational savings achieved by the proposed Tucker decomposition method. In the examples below, the performance of the proposed Tucker decomposition is compared with that of SVD method [11]. Both techniques are utilized in an FMM-FFT accelerated SIE (i.e., FMM-FFT-SIE) simulator implemented in Matlab 2019b. The simulator is executed on an Intel Xeon Gold 6142 CPU with 384 GB RAM. Unless stated otherwise, FMM box size and FMM accuracy are set to $2\lambda$ and 5 digits, respectively, while the tolerance for Tucker decomposition and SVD methods is $\gamma = 10^{-6}$. The compression ratio quantifies the memory saving achieved by the methods. It is defined as the ratio of the memory requirement of the original tensor/matrix to that of Tucker/SVD-compressed representation. Furthermore, speedup quantifies the computational saving achieved by the methods in aggregation/disaggregation stages. It is defined as the ratio of the CPU time required by the traditional matrix-vector multiplication to the CPU time required by fast tensor/matrix vector multiplications performed via Tucker/SVD-compressed representation, respectively.

### A. The Trihedral Corner Reflector

In the first example, a PEC trihedral corner reflector with three mutually perpendicular square plates with an edge length of $6\lambda$ is considered [Fig. 4(a)]. The structure is discretized by 42,477 RWG basis functions and excited by a plane-wave propagating along $\hat{\mathbf{k}}_{inc} = \sin\theta_{inc}\cos\phi_{inc}\hat{\mathbf{x}} + \sin\theta_{inc}\sin\phi_{inc}\hat{\mathbf{y}} + \cos\theta_{inc}\hat{\mathbf{z}}$ at 3 GHz, where $\theta_{inc} = 135^{\circ}$ and $\phi_{inc} = 225^{\circ}$. The incident electric field has only theta component.

*Performance and accuracy:* The Tucker and SVD enhanced FMM-FFT-SIE simulators are used to perform the analysis of EM scattering from the square trihedral corner reflector. For this analysis, the simulators are executed by setting the tolerances $\gamma = \{10^{-4}, 10^{-5}, 10^{-6}\}$. For each tolerance value, the memory requirement of the far-fields and the CPU time requirement of the aggregation and disaggregation stages are tabulated in Table I; these results are compared with those of the original FMM-FFT-SIE simulator with no decomposition enhancement. As seen in Table I, the Tucker decomposition reduces the memory requirement of far-fields by a factor of 11.5 while the SVD can reduce it by a factor of 7.5 for $\gamma = 10^{-6}$. For the same tolerance, the Tucker decomposition reduces the CPU time requirements of the aggregation and disaggregation stages by factors of 18.9 and 17.9, while the SVD method reduces the CPU time requirements of those stages by factors of 11.3 and 12.3, respectively. Apparently, the Tucker decomposition outperforms the SVD method. In Table I, the relative $L^2$−norm differences between the bistatic RCSs obtained by the original FMM-FFT-SIE simulator (with no decomposition enhancement) and Tucker and SVD enhanced FMM-FFT-SIE simulators are compared. Needless to say, the Tucker decomposition yields RCSs with much higher accuracy for $\gamma = 10^{-5}$ and $\gamma = 10^{-6}$. The RCS obtained by Tucker and SVD enhanced FMM-FFT-SIE simulators for different tolerances are also plotted and compared with the RCS obtained by the commercial software FEKO [Fig. 4]. Note that the relative $L^2$−norm difference between RCSs obtained by the original FMM-FFT-SIE simulator and FEKO is 1.13%. It is clear in Fig. 4 that the Tucker-enhanced FMM-FFT-SIE simulator provides accurate RCS for $\gamma = 10^{-5}$. Note that the RCS obtained for $\gamma = 10^{-6}$ is not included in Fig. 4 as it is identical to the RCS obtained via commercial simulator.

*Decomposition tolerance and FMM accuracy:* The performances of the Tucker and SVD methods are examined while changing the decomposition tolerance $\gamma$ from $10^{-3}$ to $10^{-7}$ [Fig. 5]. For the tolerances $10^{-3}$ and $10^{-7}$, the compression ratios achieved by the Tucker decomposition are 23 and 9.5, while those achieved by SVD method are 13.5 and 6.5, respectively [Fig. 5(a)]. For $\gamma = 10^{-3}$ and $\gamma = 10^{-7}$, the speedups in aggregation/disaggregation stages achieved by the Tucker decomposition are 34.0/32.5 and 14.2/14.9, while those achieved by the SVD method are 22.7/21.0 and 9.2/10.2, respectively. Clearly, both methods yield more compression and speedup with increasing tolerance, but Tucker performs the best [Fig. 5(b)].

TABLE I
CPU TIME AND MEMORY COST OF TUCKER AND SVD METHODS FOR THE TRIHEDRAL CORNER REFLECTOR EXAMPLE

| Decomposition Methodology | | MEMORY (MB) | AGG. TIME (S) | DISAGG. TIME (S) | REL. $L^2$−NORM DIFF. (%) |
|---|---|---|---|---|---|
| No decomposition | | 3694 | 0.340 | 0.357 | - |
| SVD | $\gamma = 10^{-4}$ | 344 | 0.019 | 0.020 | 1.48 |
| | $\gamma = 10^{-5}$ | 415 | 0.026 | 0.027 | 1.47 |
| | $\gamma = 10^{-6}$ | 493 | 0.030 | 0.029 | 0.11 |
| Tucker | $\gamma = 10^{-4}$ | 208 | 0.012 | 0.013 | 4.93 |
| | $\gamma = 10^{-5}$ | 263 | 0.017 | 0.018 | 0.65 |
| | $\gamma = 10^{-6}$ | 320 | 0.018 | 0.020 | 0.06 |

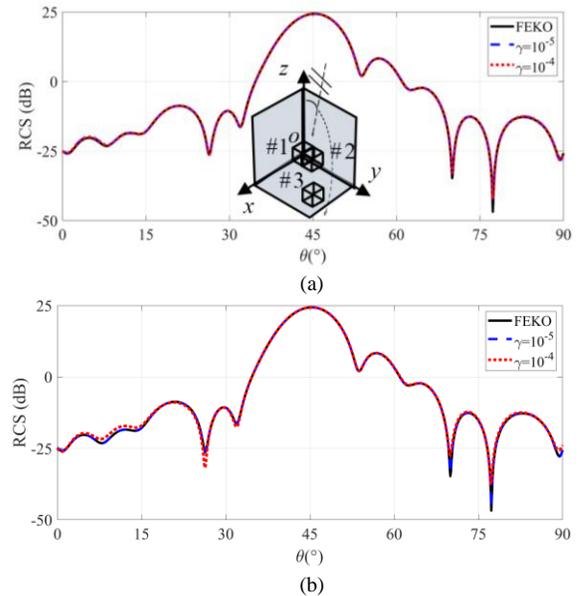

(a)

(b)

Fig. 4. Bistatic RCSs of the trihedral corner reflector at 3GHz obtained by FEKO and (a) SVD and (b) Tucker enhanced FMM-FFT-SIE simulators with $\gamma = \{10^{-4}, 10^{-5}\}$.

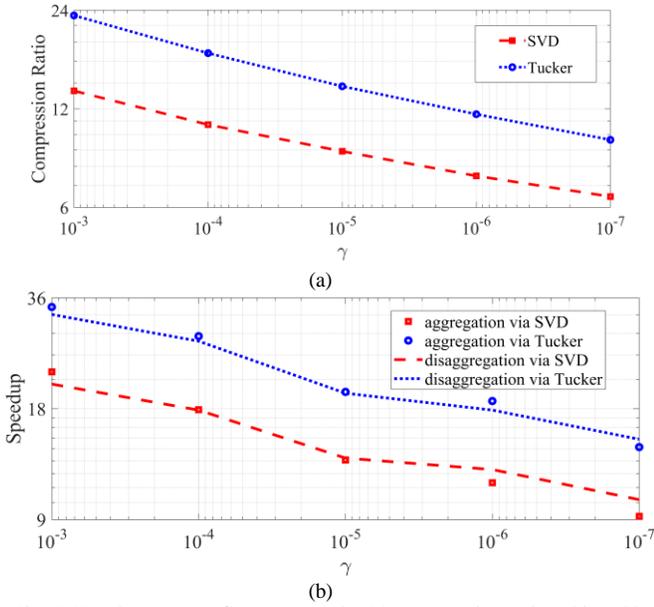

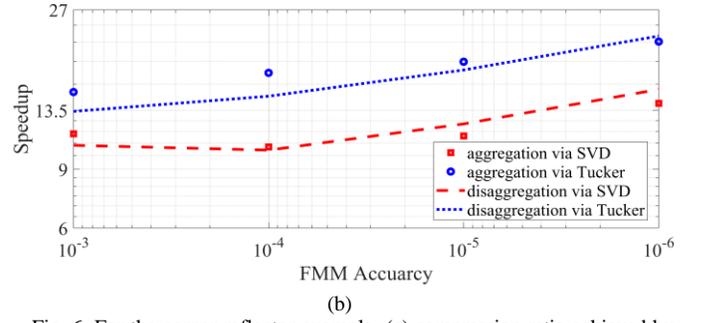

Fig. 6. For the corner reflector example, (a) compression ratio achieved by the Tucker and SVD methods and (b) the speedup achieved by the Tucker and SVD methods in aggregation and disaggregation stages while the FMM accuracy changes from $10^{-3}$ to $10^{-6}$.

*Box size and ranks:* Next, the performances of the Tucker and SVD methods are examined while the FMM box size is changed. For the box sizes $0.5\lambda$, $1\lambda$, $1.5\lambda$, and $2.0\lambda$, the average number of RWG basis functions per FMM box $N_{ave}$ is 106, 462, 1137, and 2215, respectively. For each box size, the compression ratio and speed up achieved by Tucker and SVD methods are plotted [Fig. 7]. As shown in Fig. 7(a), the compression ratios achieved by Tucker and SVD methods are 4 and 2 for box size of $0.5\lambda$ or $N_{ave}=106$; those are 11.5 and 7.5 for box size of $2\lambda$ or $N_{ave}=2215$, respectively Clearly, the compression ratio achieved by methods increases with increasing box size for Tucker and SVD methods. As shown in Fig. 7(b), the speedups achieved by Tucker and SVD methods are 2.2/1.9 and 2.4/2.2 for box size of $0.5\lambda$ or $N_{ave}=106$; those are 18.9/17.9 and 11.3/12.3 for box size of $2\lambda$ or $N_{ave}=2215$, respectively. It should be mentioned here when the box size is $0.5\lambda$, the speedup achieved by Tucker is not better than that achieved by SVD. This is because the computational cost of memory access, transpose, and reshape operations of Tucker (outlined in Figs. 2 and 3) dominates the computational time of aggregation/disaggregation stages for small matrices. On the other hand, the SVD method is free from this additional computational cost while it only requires one-time memory access. Not surprisingly, the speedup increases with increasing box size and compression ratio (and reducing numbers of elements in the representations while performing aggregation/disaggregation stages).

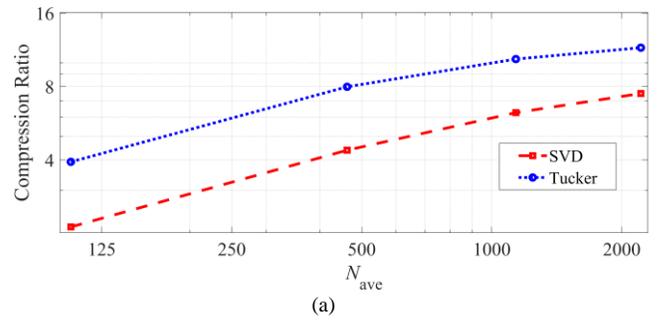

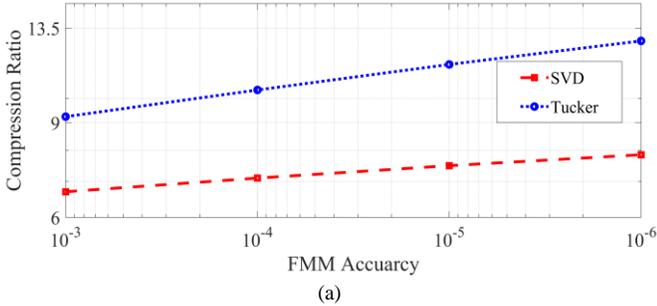

Fig. 5. For the corner reflector example, (a) compression ratio achieved by the Tucker and SVD methods and (b) the speedup achieved by the Tucker and SVD methods in aggregation and disaggregation stages while the decomposition tolerance changes from $10^{-3}$ to $10^{-7}$.

Next, the performances of the Tucker and SVD methods are demonstrated for changing FMM accuracy from $10^{-3}$ to $10^{-6}$ [Fig. 6]. For the accuracies $10^{-3}$ and $10^{-6}$, the Tucker method yields a compression ratio of 9 and 13, while the SVD method yields that of 6.7 and 7.8, respectively [Fig. 6a)]. Increasing compression ratio with increasing accuracy is not surprising as the increasing accuracy necessitates increasing $N_{dir}$ and memory requirement of far-fields. With increasing tensor/matrix size, both methods yield more compression, yet the Tucker compresses more. For the accuracies $10^{-3}$ and $10^{-6}$, the speedups in the aggregation/disaggregation stages achieved by the Tucker decomposition are 15.3/13.4 and 21.7/22.6, respectively, while those achieved by SVD method are 11.5/10.6 and 14.2/15.6, respectively [Fig. 6(b)].

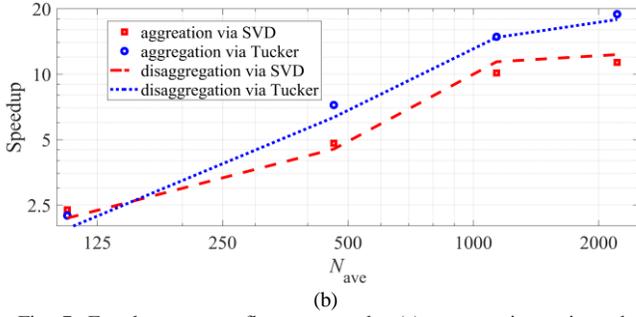

Fig. 7. For the corner reflector example, (a) compression ratio and (b) speedup achieved by Tucker and SVD methods while the average number of RWG basis functions per FMM box $N_{ave}$ increases.

Three representative FMM boxes labeled as #1, #2, and #3 in the inlet of Fig. 4(a) (from the trihedral, dihedral, and flat portion of the surface) are chosen to examine the ranks. The SVD and Tucker ranks for the box sizes $0.5\lambda$ and $2.0\lambda$ are tabulated in Table II for this three-box configuration. The compression ratio, which can tentatively be deduced from the ratio of $n_\beta$ to the $r_\beta$ for the SVD method, increases with increasing number of RWG basis functions in the boxes. It should be noted that the $r_\beta$ ranks of the SVD and Tucker methods are nearly the same [Table II]. (Note: a slight difference between the $r_\beta$ ranks of SVD and Tucker is expected as SVD and Tucker methods use $\gamma$ and $\gamma / \sqrt{3}$ for truncation, respectively.) The $r_\theta$ of Tucker is the same for all boxes (#1, #2, and #3) and only increases when the number of plane-wave directions increases. The $r_\varphi$ of Tucker slightly increases when the orientation of the basis functions in the box exhibit variations; it is the minimum for the flat surface box (box #3) and the maximum for the trihedral surface box (box #1). The $r_\beta$ is always much larger than $r_\theta$ and $r_\varphi$.

propagating along the $-z-$ direction at $f = 300$ MHz .

*Performance and accuracy:* The Tucker and SVD enhanced FMM-FFT-SIE simulators are used for the analysis by setting the tolerances $\gamma = \{10^{-4}, 10^{-5}, 10^{-6}\}$ . For each $\gamma$ value, the memory requirement of the far-fields and the CPU time requirement of the aggregation and disaggregation stages are recorded [Table III]; The recorded results are compared with those of the original FMM-FFT-SIE simulator with no decomposition enhancement. As seen in Table III, the Tucker decomposition and SVD reduce the memory requirement of far-fields by a factor of 7.8 and 3.9 for $\gamma = 10^{-6}$ , respectively. For the same tolerance, the Tucker decomposition reduces the CPU time requirements of the aggregation and disaggregation stages by factors of 15.2 and 15.8, while the SVD method reduces the CPU time requirements of those stages by factors of 8.3 and 8.6, respectively. Again, the Tucker decomposition performs better than the SVD method. In Table III, the relative $L^2$ − norm differences between the bistatic RCSs obtained by the original FMM-FFT-SIE simulator (with no decomposition enhancement) and Tucker and SVD enhanced FMM-FFT-SIE simulators are also compared. Apparently, Tucker decomposition yields accurate RCS for $\gamma = 10^{-5}$ and $\gamma = 10^{-6}$ . The RCS obtained by Tucker and SVD enhanced FMM-FFT-SIE simulators for different tolerances are compared with the RCS obtained by the Mie series solution [Fig. 8]. Note that the relative $L^2$ − norm difference between RCSs obtained by the original FMM-FFT-SIE simulator and Mie series solution is 0.01%. It is clear in Fig.8 that the Tucker-enhanced FMM-FFT-SIE simulator provides accurate RCS for $\gamma = 10^{-5}$ . Note that the RCS for $\gamma = 10^{-6}$ is not plotted in Fig. 8 since it is indistinguishable from the analytical solution.

TABLE II
RANKS OF TUCKER AND SVD METHODS FOR DIFFERENT BOX SIZES IN THE CORNER REFLECTOR EXAMPLE

| | Compression Scheme | SVD | TUCKER | | |
| Box Size | | $r_\beta$ | $r_\theta$ | $r_\varphi$ | $r_\beta$ |
|---|---|---|---|---|---|
| $0.5\lambda$ ($L=16$) | #1 ($n_\beta = 283$) | 90 | 16 | 20 | 94 |
| | #2 ($n_\beta = 190$) | 75 | 16 | 18 | 79 |
| | #3 ($n_\beta = 99$) | 45 | 16 | 16 | 48 |
| $2\lambda$ ($L=37$) | #1 ($n_\beta = 4725$) | 292 | 27 | 38 | 304 |
| | #2 ($n_\beta = 3149$) | 233 | 27 | 34 | 238 |
| | #3 ($n_\beta = 1564$) | 134 | 27 | 29 | 138 |

TABLE III
CPU TIME AND MEMORY COST OF TUCKER AND SVD METHODS FOR THE SPHERE EXAMPLE

| Decomposition Methodology | | MEMORY (MB) | AGG. TIME (S) | DISAGG. TIME (S) | REL. $L^2$ − NORM DIFF. (%) |
|---|---|---|---|---|---|
| No decomposition | | 81784 | 7.577 | 8.212 | - |
| SVD | $\gamma = 10^{-4}$ | 13665 | 0.648 | 0.705 | 0.042 |
| | $\gamma = 10^{-5}$ | 17368 | 0.944 | 0.953 | 0.004 |
| | $\gamma = 10^{-6}$ | 20914 | 0.917 | 0.935 | 0.002 |
| Tucker | $\gamma = 10^{-4}$ | 6153 | 0.328 | 0.362 | 0.540 |
| | $\gamma = 10^{-5}$ | 8230 | 0.492 | 0.520 | 0.040 |
| | $\gamma = 10^{-6}$ | 10529 | 0.498 | 0.519 | 0.002 |

### B. The Sphere

In the second example, the Tucker decomposition's performance is demonstrated through its application to the analysis of EM scattering from a $30\lambda$ -diameter sphere centered at the origin. The PEC sphere is discretized by 940308 basis functions and excited by a $x-$ polarized plane-wave

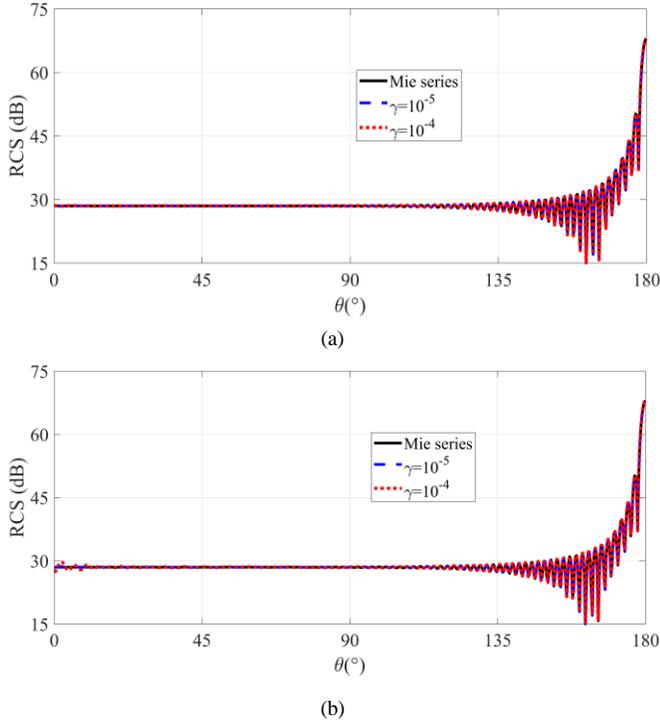

Fig. 8. Bistatic RCSs of the PEC sphere at 300MHz obtained by the analytical Mie series solution and the (a) SVD and (b) Tucker enhanced FMM-FFT-SIE solver with $\gamma = \{10^{-4}, 10^{-5}\}$ .

*Decomposition tolerance and FMM accuracy:* The performances of Tucker and SVD methods are tested while the decomposition tolerance $\gamma$ is varied from $10^{-3}$ to $10^{-7}$ [Fig. 9]. For $\gamma = 10^{-3}$ and $\gamma = 10^{-7}$, the compression ratios achieved by the Tucker decomposition are 19 and 6, while those achieved by SVD method are 8 and 3, respectively [Fig. 9(a)]. For $\gamma = 10^{-3}$ and $\gamma = 10^{-7}$ , the speedups in aggregation/disaggregation stages achieved by the Tucker decomposition are 25.6/25.3 and 10.2/10.4, while those achieved by the SVD method are 13.2/14.4 and 5.3/6.1, respectively [Fig. 9(b)]. Again, both methods yield more compression and speedup with increasing tolerance.

Next, the FMM accuracy is varied from $10^{-3}$ to $10^{-6}$ [Fig. 10]. For the accuracies $10^{-3}$ and $10^{-6}$ , the Tucker method yields a compression ratio of 6 and 8.6, while the SVD method yields that of 3.6 and 4, respectively [Fig. 10(a)]. For the accuracies $10^{-3}$ and $10^{-6}$ , the speedups in the aggregation/disaggregation stages achieved by the Tucker decomposition are 9.2/9.1 and 13.9/14.9, respectively, while those achieved by SVD method are 6.2/6.5 and 6.6/7.9, respectively [Fig. 10(b)]. With increasing FMM accuracy, the number of plane-wave directions increases. As a result, compression ratio and speedup achieved by Tucker decomposition increase faster compared to those achieved by SVD method.

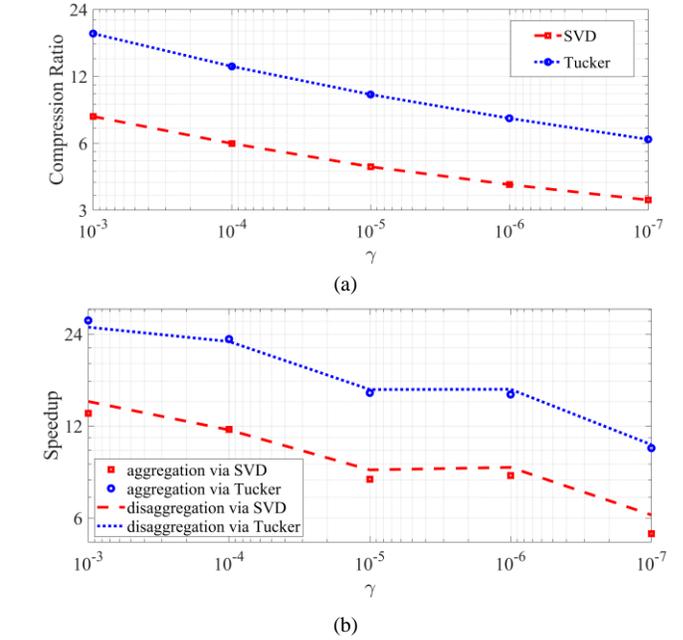

Fig. 9. For the sphere example, (a) compression ratio achieved by the Tucker and SVD methods and (b) the speedup achieved by the Tucker and SVD methods in aggregation and disaggregation stages while $\gamma$ changes from $10^{-3}$ to $10^{-7}$ .

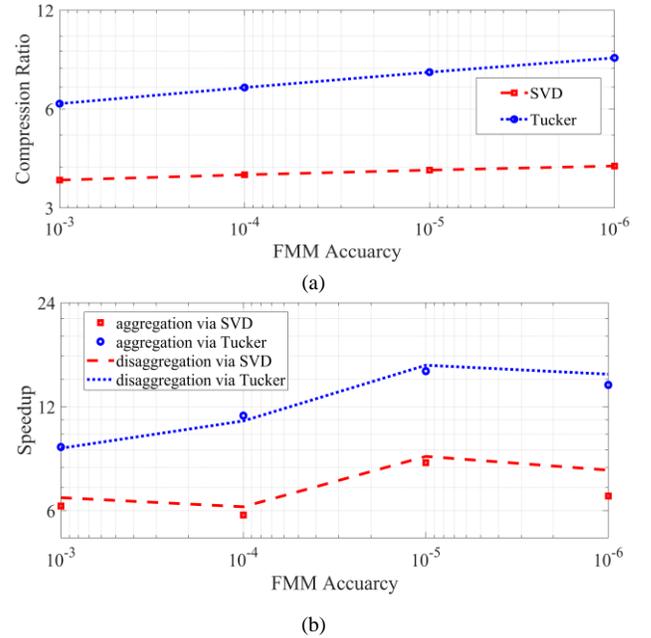

Fig. 10. For the sphere example, (a) compression ratio achieved by the Tucker and SVD methods and (b) the speedup achieved by the Tucker and SVD methods in aggregation and disaggregation stages while FMM accuracy changes from $10^{-3}$ to $10^{-6}$ .

*Box size and ranks:* Next, the performances of the Tucker and SVD methods are tested when the FMM box size is set to $0.5\lambda$ , $1\lambda$ , $1.5\lambda$ , and $2.0\lambda$ ; the average number of RWG basis functions per FMM box $N_{ave}$ for each box size is 58, 232, 513, and 909, respectively. For each box size, the compression ratio and speed up achieved by Tucker and SVD methods are plotted [Fig. 11]. As shown in Fig. 11(a), the compression ratios achieved by Tucker and SVD methods are 2 and 1.2 for box size of $0.5\lambda$ or $N_{ave} = 58$; those are 7.8 and 3.9 for box size of $2\lambda$ or $N_{ave} = 909$, respectively. Clearly, the compression ratio achieved by methods increases with increasing box size for Tucker and SVD methods. As shown in Fig. 11(b), the speedups in aggregation/disaggregation stages achieved by Tucker and SVD methods are 1.2/1.1 and 1.3/1.1 for box size of $0.5\lambda$ or $N_{ave} = 58$; those are 15.2/15.8 and 8.3/8.6 for box size of $2\lambda$ or $N_{ave} = 909$, respectively. Again, the speedup achieved by Tucker is not as good as the memory saving it achieves for $0.5\lambda$ box size. This is because of the domination of the computational cost of the matrix transpose and reshape operations (outlined in Figs. 2 and 3) and multiple-times memory access for the operations with small matrices. Furthermore, the speedup increases with increasing box size and compression ratio. These results show that the Tucker is better than SVD for flat surfaces (as in the trihedral example) and curved surfaces (as in sphere).

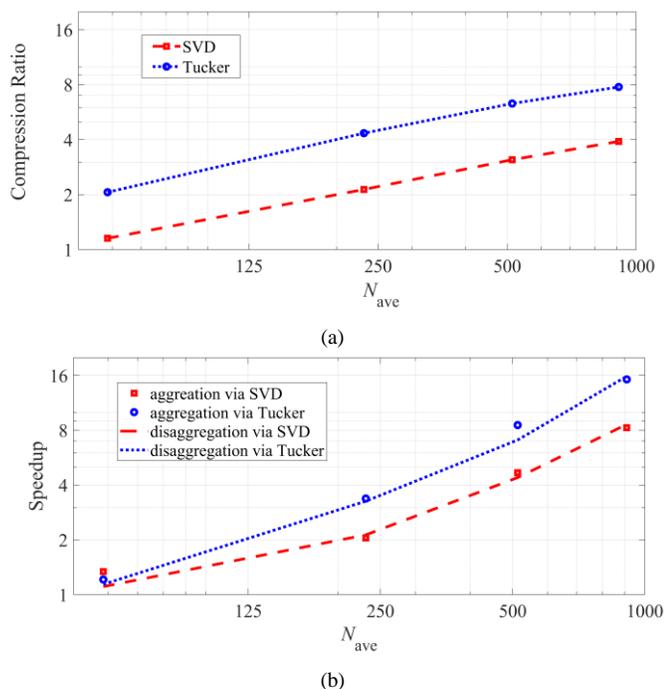

Fig. 11. For the sphere example, (a) compression ratio and (b) speedup achieved by Tucker and SVD methods while the average number of RWG basis functions per FMM box increases.

TABLE IV
RANKS OF TUCKER AND SVD METHODS FOR DIFFERENT BOX SIZES IN THE SPHERE EXAMPLE

| Compression Scheme / Box Size | SVD | | TUCKER | |
|---|---|---|---|---|
| | $r_\beta$ | $r_\theta$ | $r_\varphi$ | $r_\beta$ |
| $0.5\ \lambda$   ($L=16, n_\beta=58$) | 46 | 15 | 17 | 47 |
| $1\ \lambda$   ($L=24, n_\beta=232$) | 89 | 18 | 22 | 93 |
| $1.5\ \lambda$   ($L=31, n_\beta=513$) | 127 | 22 | 25 | 132 |
| $2\ \lambda$   ($L=37, n_\beta=909$) | 168 | 25 | 29 | 175 |

Finally, Tucker and SVD ranks are listed for different box sizes [Table IV]. The ratio of $n_\beta$ to the $r_\beta$ for the Tucker and SVD methods increases with increasing box size and $N_{ave}$. Furthermore, $r_\theta$ and $r_\varphi$ of Tucker slightly increases with increasing box size (or the number of plane-wave directions). For all box sizes, $r_\beta$ is always much larger than $r_\theta$ and $r_\varphi$.

## IV. CONCLUSION

In this paper, a Tucker decomposition methodology was proposed to compress aggregation and disaggregation matrices in the FMM-accelerated SIE simulators. The proposed Tucker enhancement yields a significant reduction in the memory requirement of these matrices as well as the computational cost of the aggregation and disaggregation stages. It can be easily incorporated into the existing FMM-accelerated SIE simulators. The proposed Tucker scheme yields better compression ratios and speedups in the aggregation and disaggregation stages compared to the SVD. The proposed methodology's performance was extensively tested for varying simulation parameters such as decomposition tolerance, FMM accuracy, and FMM box size.

## REFERENCES

[1] Ö. Ergül, "Analysis of composite nanoparticles with surface integral equations and the multilevel fast multipole algorithm," *J. Opt.*, vol. 14, no. 6, pp. 062701-1–062701-4, 2012.

[2] B. Karaosmanoglu and Ö. Ergül, "Modified combined tangential formulation for stable and accurate analysis of plasmonic structures," *ACES Exp.*, vol. 34, no. 5, pp. 811–814, 2019.

[3] A. C. Yucel, W. Sheng, C. Zhou, Y. Liu, H. Bağcı, and E. Michielssen, "An FMM-FFT accelerated SIE simulator for analyzing EM wave propagation in mine environments loaded with conductors," *IEEE Journal on Multiscale and Multiphysics Comp. Techn.*, vol.3, pp. 3-15, 2018.

[4] A. C. Yucel, Y. Liu, H. Bağcı, and E. Michielssen, "Statistical characterization of electromagnetic wave propagation in mine environments," *IEEE Antennas Wireless Propag. Lett.*, vol. 12, pp. 1602-1605, 2013.

[5] I. I. Giannakopoulos, M. S. Litsarev, and A. G. Polimeridis, "Memory footprint reduction for the FFT-based volume integral equation method via tensor decompositions," *IEEE Trans. Antennas Propagat.*, vol. 67, no. 12, pp. 7476-7486, Dec. 2019.

[6] Z. Chen, S. Zheng, and V. I. Okhmatovski, "Tensor train accelerated solution of volume integral equation for 2-D scattering problems and magneto-quasi-static characterization of multiconductor transmission lines," *IEEE Trans. Microw. Theory Tech.*, vol. 67, no. 6, pp. 2181-2196, 2019.

[7] C. Qian and A. C. Yucel, "On the compression of translation operator tensors in FMM-FFT-Accelerated SIE simulators via tensor decompositions," *IEEE Trans. Antennas Propag.*, early access, 2020.

[8] A. C. Yucel, L. J. Gomez, and E. Michielssen, "Compression of translation operator tensors in FMM-FFT accelerated SIE solvers via Tucker decomposition," *IEEE Antennas Wireless Propag. Lett.*, vol. 16, pp. 2667-2670, 2017.

[9] C. Qian, Z. Chen, and A. C. Yucel, "Tensor decompositions for reducing the memory requirement of translation operator tensors in FMM-FFT accelerated IE solvers," in *Proc Applied Computational EM Society (ACES) Symp.*, Apr. 2019, pp. 1-2.

[10] J. Wei, Z. Peng, and J. F. Lee, "Multiscale electromagnetic computations using a hierarchical multilevel fast multipole algorithm," *Radio Science*, vol. 49, no. 11, pp.1022-1040, 2014.

[11] J. L. Rodriguez, J. M. Taboada, M. G. Araujo, F. Obelleiro Basteiro, L. Landesa, and J. I. Garcia-Tunon, "On the use of the singular value decomposition in the fast multipole method," *IEEE Trans. Antennas Propagat.*, vol. 56, no. 8, pp. 2325-2334, 2008.

[12] S. He, Z. Nie, and J. Hu, "Electromagnetic solution for dielectric objects with multilevel fast multipole algorithm and singular value decomposition," in *Proc IEEE Antennas and Propagation Society International Symposium*, Jun. 2009, pp. 1-4.

[13] A. Anandkumar, P. Jain, Y. Shi, and U. N. Niranjan, "Tensor vs. matrix methods: Robust tensor decomposition under block sparse perturbations," *Artificial Intelligence and Statistics*, vol. 51, pp. 268-276, 2016.

[14] N. B. Erichson, K. Manohar, S. L. Brunton, and J. N. Kutz, "Randomized CP tensor decomposition," *Machine Learning: Sci. and Techn.*, vol. 1, no. 2, p. 025012, 2020.

[15] M. Wang, C. Qian, and A. C. Yucel, "Tucker-enhanced VoxCap simulator for electrostatic analysis of voxelized structures," in *Proc. IEEE MTT-S Int. Conf. on Num. EM and Multiphysics Modeling and Opt. (NEMO)*, Cambridge, MA, USA, May 2019, p. 1.

[16] M. Wang, C. Qian, Zhuotong Chen, E. D. Lorenzo, L. J. Gomez, S. Zheng, V. Okhamatovski, and A. C. Yucel, "Tucker-enhanced VoxHenry simulator for inductance extraction of voxelized conducting/superconducting structures," in *IEEE MTT-S Int. Conf. on Num. EM and Multiphysics Modeling and Opt. (NEMO)*, Cambridge, MA, USA, May 2019, p. 1.

[17] M. Wang, C. Qian, J. K. White, and A. C. Yucel, "VoxCap: FFT-accelerated and Tucker-enhanced capacitance extraction simulator for voxelized structures," *IEEE Trans. Microw. Theory Tech,.* early access, 2020.

[18] Z. Chen, S. Zheng, Q. Cheng, A. C. Yucel, and V. Okhmatovski, "Pre-corrected tensor train algorithm for current flow modelling in 2D multi-conductor transmission lines," in *2019 IEEE MTT-S International Microwave Symposium (IMS)*, Jun. 2019, pp. 124-127.

[19] Z. Chen, L. G. Gomez, S. Zheng, A. C. Yucel, and V. Okhmatovski, "Sparsity aware pre-corrected tensor train algorithm for fast solution of 2D scattering problems and current flow modelling on unstructured meshes," *IEEE Trans. Microw. Theory Tech.*, vol. 67, no. 12, pp. 4833-4847, 2019.

[20] S. M. Rao, D. R. Wilton, and A. W. Glisson, "Electromagnetic scattering by surfaces of arbitrary shape," *IEEE Trans. Antennas Propagat.*, vol. 30, no. 3, pp. 409-418, May 1982.

[21] R. Coifman, V. Rokhlin, and S. Wandzura, "The fast multipole method for the wave equation: a pedestrian prescription," *IEEE Antennas Propagat. Mag.*, vol. 35, no. 3, pp. 7-12, 1993.

[22] R. L. Wagner, S. Jiming, and W. C. Chew, "Monte Carlo simulation of electromagnetic scattering from two-dimensional random rough surfaces," *IEEE Trans. Antennas Propagat.*, vol. 45, no. 2, pp. 235-245, 1997.

[23] J. Song, C. C. Lu, and W. C. Chew, "Multilevel fast multipole algorithm for electromagnetic scattering by large complex objects," *IEEE Trans. Antennas Propagat.*, vol. 45, no. 10, pp. 1488-1493, 1997.

[24] A. C. Yucel, "Helmholtz and high-frequency Maxwell multilevel fast multipole algorithms with self-tuning library," MSc dissertation, Electrical Engineering and Computer Science, Univ. Michigan, Ann Arbor, MI, USA, 2008.

[25] W. C. Chew, E. Michielssen, J. M. Song, and J. M. Jin, *Fast and efficient algorithms in computational electromagnetics*. Norwood, MA, USA: Artech House, Inc., 2001.

[26] L. D. Lathauwer, B. D. Moor, and J. Vandewalle, "A multilinear singular value decomposition," *SIAM J. Matrix Anal. Appl.*, vol. 21, no. 4, pp. 1253-1278, 2000.

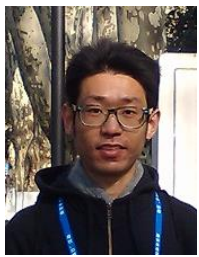

**Cheng Qian** received the B.S. and Ph.D. degree in electronics engineering from Nanjing University of Science and Technology, Jiangsu, China, in 2009 and 2015. From 2016 to 2018, he was a Research Associate with the Department of Applied Physics, The Hong Kong Polytechnic University, Hong Kong. Since 2019, he has been a Post-Doctoral Researcher with the School of Electrical and Electronic Engineering, Nanyang Technological University, Singapore.

His current research interests include computational electromagnetics and nonlinear plasmonics.

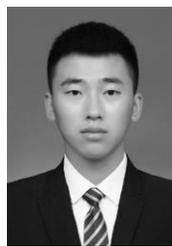

**Mingyu Wang** received the B.S. degree in electrical engineering and automation from Northeast Forestry University, China, in 2016 and M.S. in electronics from Nanyang Technological University, Singapore, in 2018. He is currently working toward the Ph.D. degree at the School of Electrical and Electronic Engineering, Nanyang Technological University, Singapore.

His research is focused on computational electromagnetics and fast parameter extraction of voxelized interconnects and circuits.

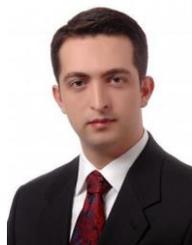

**Abdulkadir C. Yucel** (M'19-SM'20) received the B.S. degree in electronics engineering (*Summa Cum Laude*) from Gebze Institute of Technology, Kocaeli, Turkey, in 2005, and the M.S. and Ph.D. degrees in electrical engineering from the University of Michigan, Ann Arbor, MI, USA, in 2008 and 2013, respectively.

From August 2006 to April 2013, he was a Graduate Student Research Assistant at the University of Michigan. Between May 2013 and December 2017, he worked as a Postdoctoral Research Fellow in various institutes, including the Massachusetts Institute of Technology. Since 2018, he has been working as an Assistant Professor at the School of Electrical and Electronic Engineering, Nanyang Technological University, Singapore.

Dr. Yucel received the Fulbright Fellowship in 2006, Electrical Engineering and Computer Science Departmental Fellowship of the University of Michigan in 2007, and Student Paper Competition Honorable Mention Award at IEEE AP-S in 2009. He has been serving as an Associate Editor for the International Journal of Numerical Modelling: Electronic Networks, Devices and Fields and as a reviewer for various technical journals. His research interests include various aspects of computational electromagnetics with emphasis on analytical and numerical electromagnetic modelling and the applications of uncertainty quantification and deep/machine learning techniques to the electromagnetic analyses.